\documentclass[10pt,preprint]{aastex}
\usepackage{graphicx}

\begin{document}

\noindent

{\bf QUASARS, GAMMA RAY BURSTERS AND BL LACERTIDS}

Halton Arp 

\noindent {\bf New observations suggest that high redshift quasars can
be turned into Gamma Ray Bursters and BL Lacertids by interaction with
absorbing clouds in their vicinity.}

 Like quasars, Gamma Ray Bursters (GRB's) are high redshift objects
which emit copious amounts of high energy radiation in their outburst
phases. Recently a startling observation for which experts have
no plausible explanation was reported $^{1,2}$. The new evidence
shows that supposedly intervening galaxies are 4 times more prevalent
along lines of sight to GRB's than to quasars. 
Since quasars and GRB's of the same redshift are supposed to be at the
same extraordinarily large distances we are sampling a given path
length through the Universe in different directions. To put it most
simply: {\it If the only difference between these path lengths is that
GRB path lengths have more absorbing clouds, then these Gamma
Ray Bursters and the absorbing clouds must be physically associated.}

Because GRB's are very active objects they would be expected to have
disrupted components and ejected clouds in their vicinity. One obvious
possibility is that these furnish the the intervening absorption lines
which are seen in their spectra. Clouds ejected in the direction of the
observer would have smaller Doppler shifts.

The observed shifts in the absorption lines, however, can approach the
speed of light and are in general much too high to be dispersion
velocities in a group of extragalactic objects. Also the absorption
systems are not generally excited and seem to represent fairly
quiescent clouds.

An alternative suggestion is that the absorbing clouds are older
ejections and remmnants which had intrinsic redshifts which have now
partially decayed with time. If we rule out high approach
velocities we observationally have a physical association of high and
low redshift objects which are supposedly at much differerent
distances. Where have we seen that before? Simply for 40 years in the 
association of high redshift quasars with lower redshift galaxies 
$^{3,4}$. Since this appears to be the only alternative it is
interesting to examine it in a little more detail. 

What we seem to be seeing is high intrinsic redshift GRB's at the closer
distance of the absorbing clouds. Spatially the GRB's must be in
closer proximity to their associated galaxies (or clouds of gaseous
material ripped out in the process of explosive
events). The difference between the quasars, which have also been shown
to be much closer than their redshifts conventionally dictate, and the
GRB's would then be that the GRB's tend to burst more
violently, perhaps carrying older gas clouds from their parent
galaxies with them or in their wake. Or simply encountering them in
the extended vicinity. The quasars on the other hand would come out
more cleanly, for example, along the minor axes of their ejecting
galaxies$^5$. (The latter reference being very direct obervational
confirmation for ejection of intrinsically redshifted quasars.)

In this case the whole ensemble of quasars and GRB's with their
galaxies/gas are much closer to the observer than their redshifts
would conventionally place them. We would then have nearby clusters
and groups with a range of object types. They would have a mixture of
different ages and differing intrinsic redshifts. This is what the
observations seem to require in galaxy/quasar associations.

As further support of this interpretation one can cite Stocke and
Rector in 1997$^6$ that BL Lac objects also have excess MG II
absorbers in their line of sight. This represents a different class of
high red shift objects which show the same excess of MGII absorbers as
in the GRB's. The important point is that BL Lac objects are like quasars
but with their gaseous outer layers stripped away and only the
continuum emitting surface spectroscopically observable. What could
cause this difference between quasars and BL Lac's? The obvious answer
would be that the latter was the result of a collision with a gas cloud and
thus gave rise to another variety of quasar similar to the case of the
GRB's. As further support for this interpretation the BL Lacs are
observed to be found closer in angular separation from active parent
galaxies$^7$). (This was found from Ultra Luminous X-ray sources which
were accepted as belonging to low redshift galaxies. The objects
turned out to be essentially all high redshift quasars but with a
higher percentage of BL Lac's. Presumably the stripping collision
stopped the BL Lacs closer to the parent galaxy.)             

A key object extensively observed by Margaret Burbidge is 
AO 0235 +164$^8$. It seems to unlock in detail the absorption excesses 
in front of the GRB's and BL Lac's which has puzzled the established
explanations. 

Surely these different redshift objects within a few arcsec of each
other are at the same distance. The high redshift BL Lac has z= .940.
Absorptions take place in or closely behind clouds emitted by the z =
.524 galaxy and a z = .851 system. The fact that the emission in the
the BL Lac at z = .940 is so weak would indicate its outer emission
layers had been almost completely stripped in a collision.

The developments discussed here have now shown excess absorption
clouds are typical of sight lines to BL Lac's and GRB's. That means the
close coincidence of the AO 0235 objects is not just an extremely
unlikely accident but is established as a class characteristic.
The finishing touch is that the BL Lac AO 0235 +164 is a very strong
radio source, IR source {\bf and a Gamma Ray Source!}.

The summary conclusion then becomes a picture where a quasar is ejected
from an active galaxy, travels out into intergalactic space. If it
meets a cloud it can become a BL Lac with a strong continuum spectrum
which can show a range of weak lines - or a GRB with strong high energy
radiation. The nature of the resultant high redshift object should be
determined by the density of the original quasar and the the density
of the cloud impacted as well as the relative speed and directness of the
collision. GRB's, BL Lacs and quasars would then all be different
forms of the same kind of object.

\noindent Halton Arp\\
Max-Planck-Institut f\"ur Astrophysik,\\
D-85741 Garching, Germany\\
email:arp@mpa-garching.mpg.de

\section*{References}

\noindent 1. Prochter, Gabriel E.; Prochaska, Jason X.; Chen, Hsiao-Wen;
et al. 2006, On the Incidence of Strong Mg II Absorbers along
Gamma-Ray Burst Sight Lines, ApJ, 648L, 93

\noindent 2. Schilling, G. 2006, Do Gamma Ray Bursts Always Line Up With
Galaxies? Science 313, 749 
    
\noindent 3. Arp, H. 1998, Seeing Red: Redshifts, Cosmology and Academic
Science, Apeiron, Montreal\\
\noindent Arp, H. 2003, Catalogue of Discordant Redshift Associations,
Apeiron, Montreal
\medskip

\noindent 4. Burbidge, G. 2003, The Sources of Gamma-Ray Bursts and Their
Connections with QSOs and Active Galaxies, ApJ 585, 112B

\noindent 5. L\'opez-Corredoira and Guti\'errez C. 2006, astro-ph 0609514

\noindent 6. Stocke, J. and Rector T. 1997 An Excess of Mg II Absorbers
in BL Lacertae Objects ApJ 489L, 17

\noindent 7. Arp, H., Guti\'errez, C. M., L\'opez-Corredoira, M. 2004,
New optical spectra and general discussion on the nature of
ULX's, A\&A, 418, 877

\noindent 8. Cohen, R., Smith, H. E., Junkkarinen, V.,Burbidge, E. M. 1987
318, 577

\end{document}